\documentclass[12pt]{iopart}

\usepackage{graphicx}
\usepackage{bm}
\usepackage{latexsym}
\usepackage{amssymb}

\newcommand{\lacuvo}{La$_3$Cu$_2$VO$_9$}

\begin{document}

\title{Hierarchical geometric frustration in \lacuvo}
\author{J. Robert\dag, B. Canals\dag, V. Simonet\dag, R. Ballou\dag, C. Darie\ddag, B. Ouladdiaf\S, M. Johnson\S}
\address{\dag\ Laboratoire Louis N\'{e}el, CNRS, 25 avenue des
Martyrs, B.P. 166, 38 042 Grenoble Cedex 9, France}
\address{\ddag\ Laboratoire de Cristallographie, CNRS, 25 avenue des
Martyrs, B.P. 166, 38 042 Grenoble Cedex 9, France}
\address{\S\ Institut Laue-Langevin, BP 154, 38042 Grenoble Cedex,
France.} \ead{julien.robert@grenoble.cnrs.fr}

\begin{abstract}
The crystallographic structure and magnetic properties of the
\lacuvo\ were investigated by powder neutron diffraction and
magnetization measurements. The compound materializes geometric
frustration at two spatial scales, within clusters and between
clusters, and at different temperature scales. It is shown by
exactly solving the hamiltonian spectrum that collective spins are
formed on each clusters at low temperature before inter-clusters
coupling operates.
\end{abstract}


\section{Introduction}
The physics of geometrically frustrated magnets has attracted a lot
of attention in the recent years, because of the number of novel
phases they would generate either in toy models of statistical
physics or in condensed matter realizations
\cite{moessner,misguich}.
The field of molecular magnetism has also raised strong interest by
offering a unique playground for testing quantum physics fundaments
and simultaneously being at the forefront of applied research
\cite{Mn12QT,Berry}.
In this work, we investigate a two dimensional magnet, the layered
oxide \lacuvo, which exhibits many of the properties one may
encounters in frustrated as well as in molecular magnets.
The only magnetic ions in this compound are the Cu$^{2+}$
ions which are antiferromagnetically coupled and form planar
clusters of 4 corner-sharing triangles \cite{jansson,vander},
thus materializing geometric frustration of quantum spins 1/2
in a nanomagnet.
These clusters are themselves antiferromagnetically coupled, leading
to a novel two dimensional frustrated structure, which can be seen
as a triangular lattice of 9-spins units similarly as the kagom\'e
lattice is seen as a triangular lattice of 3-spins units (triangles)
(see fig. 6 in ref. \cite{vander}).
As explained later on, inter unit couplings are much weaker than
intra unit couplings.
This offers the unique opportunity to study geometrical
frustration with two spatial scales, themselves resolved by two
different energy scales.

We report in this paper some results of our current investigations
of the \lacuvo~ oxide \cite{robert}, concerned with its
crystallography and its magnetism, and provide with a quantitative
analysis of its magnetic properties, within the temperature where
these are answerable to independent clusters.

\section{\label{sec:Exp}Experimental}

\subsection{\label{sec:Magn}Cristallographic properties}

\begin{figure}
\begin{minipage}[b]{0.5\linewidth}
\centering
\includegraphics[bb=190 110 530 450,scale=0.4]{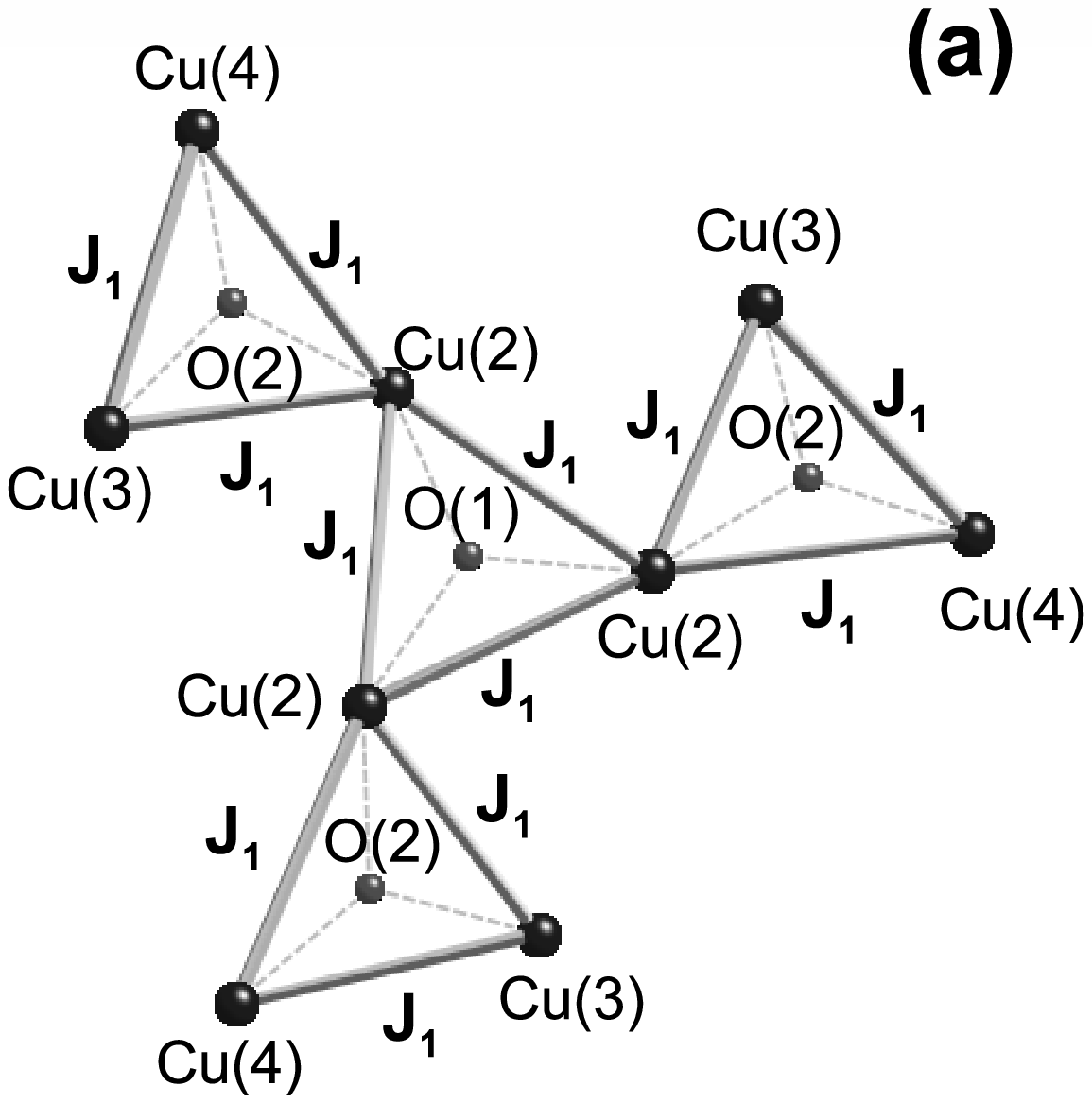}
\end{minipage}
\hspace{0.5cm} 
\begin{minipage}[b]{0.5\linewidth}
\centering
\includegraphics[bb=190 110 530 450,scale=0.4]{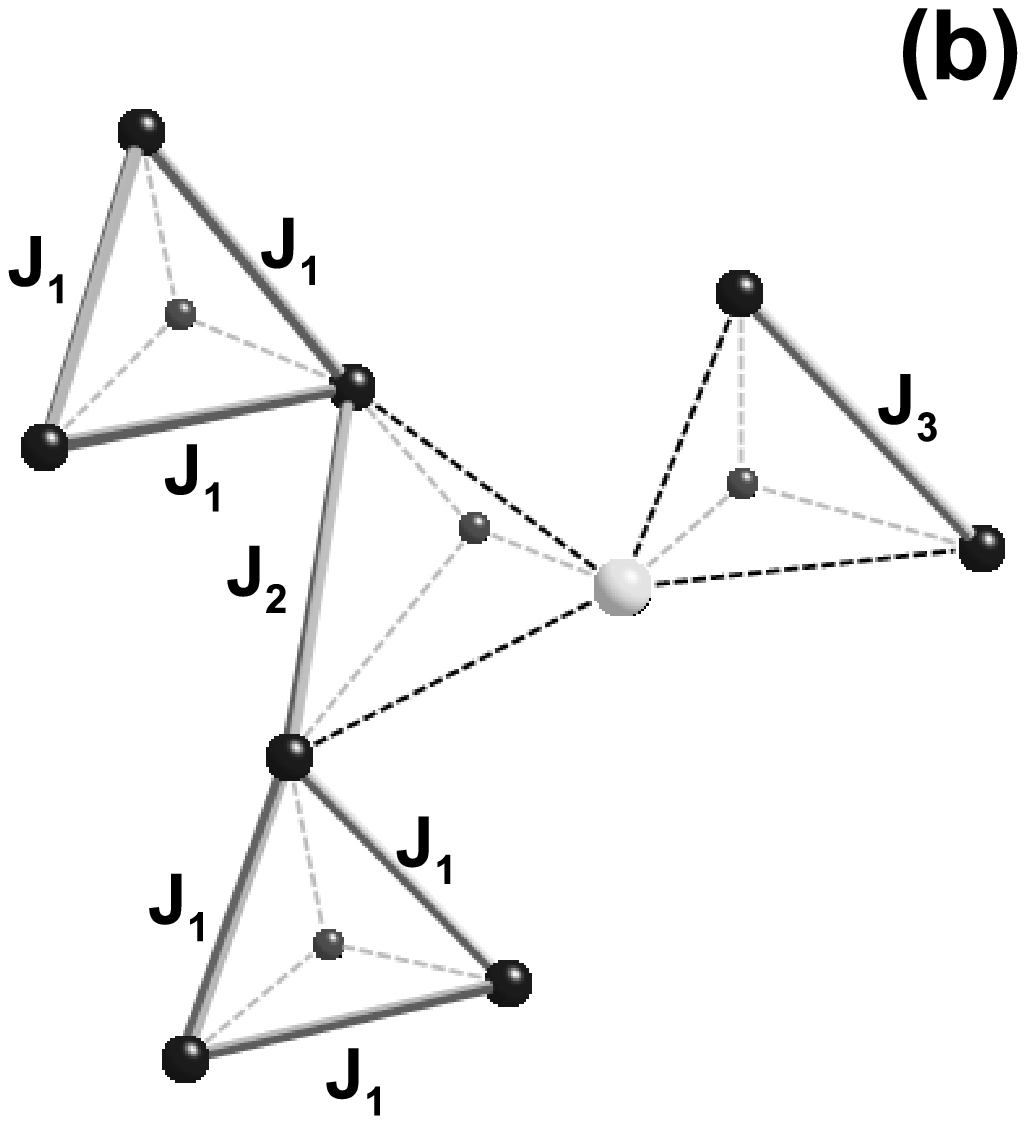}
\end{minipage}
\caption{Atomic arrangement in the 9-spins clusters (a) and
8-spins clusters (b), calculated with DFT methods, of the Cu
(large black), V (large gray) and O (small black) atoms. Labeling
of the different superexchange interactions as discussed in
section \ref{sec:model}.} \label{fig:substitutions}
\end{figure}

The \lacuvo~oxide crystallizes in the hexagonal P6$_3/m$ space group
with lattice parameters a = b = 14.395~\AA\ and c = 10.657~\AA\ at
300 K. The LaO$_{6/3}$ layers alternate with (Cu/V)O$_{3/3}$ layers
illustrating the 2D character of the structure. Within these, the
Cu$^{2+}$ ions are distributed over three inequivalent sites Cu(2),
Cu(3) and Cu(4) (following the site labeling of ref. \cite{vander})
having respectively distorted trigonal bipyramidal (Cu(3),Cu(4)) and
tetrahedral (Cu(2)) coordination environments. These form the planar
clusters of 9 spins 1/2 made of 4 corner-sharing triangles (cf. Fig.
\ref{fig:substitutions} (a)). Each cluster is centered at the vertex
of a 2D triangular lattice. On examining the crystal structure, we
observe that there are two short intralayer exchange paths
respectively mediated by one and two oxygen while the shortest
interlayer exchange is mediated by two oxygen. We would then expect
that intralayer coupling is larger than the interlayer one.

Polycrystalline \lacuvo\ samples were synthesized by a sol-gel
method. The stoichiometric metallic cations were dissolved in nitric
acid before being complexed by addition of EDTA
(ethylenediaminetetraacetic acid) in a controlled pH solution. This
solution was polymerized and then heated at 700~$^{\circ }$C to
eliminate the organic constituents. The resulting powders were
annealed during 15 days at 1010~$^{\circ }$C.
The structural quality of the samples were checked at 300 K by X-ray
analysis. The crystal structure was confirmed and no impurities were
detected.

A high resolution powder neutron diffraction experiment was
conducted on the D2B diffractometer at the Institut Laue-Langevin,
to get further insights about the crystal structure and because in
stoichiometric proportions a slight excess of vanadium ions results
in the presence of substitutions of V$^{5+}$ ions for Cu$^{2+}$
ions.
The pattern collected at 3 K was refined by allowing
V substitution on each Cu site, alternatively.
The fit is slightly better when the Cu(2) site is substituted by
approximately 9 \% of V and the Cu(3) and Cu(4) sites are
unsubstituted, which is compatible with the result obtained in
\cite{vander}. The deduced statistical population of clusters
consists then in 66\% 9-Cu clusters, 30\% 8-Cu clusters and 4\% 7-Cu
clusters, the 6-Cu clusters being practically inexistent. However,
the very weak sensitivity of the fit to the Cu$^{2+}$ site occupancy
underlines the need of a further detailed investigation.

Electronic structure calculations were performed, within the Density
Functional Theory using the VASP package \cite{vasp}, with periodic
boundary conditions and a unit cell containing 130 atoms, including
2 clusters located in the two (Cu/V)O$_{3/3}$ layers at z = 1/4 and
z = 3/4. These suggested that the 7-Cu clusters are energetically
unfavored and that the compound should in effect be considered as
made of the clusters shown in figure \ref{fig:substitutions}, in
proportion of $2/3$ for 9-Cu clusters and $1/3$ for 8-Cu clusters.
We also summarize in this figure the type of expected superexchange
interactions within the clusters.

\subsection{\label{Magn}Magnetic properties}

\begin{figure}
\begin{minipage}[b]{0.5\linewidth}
\centering
\includegraphics[scale=0.7]{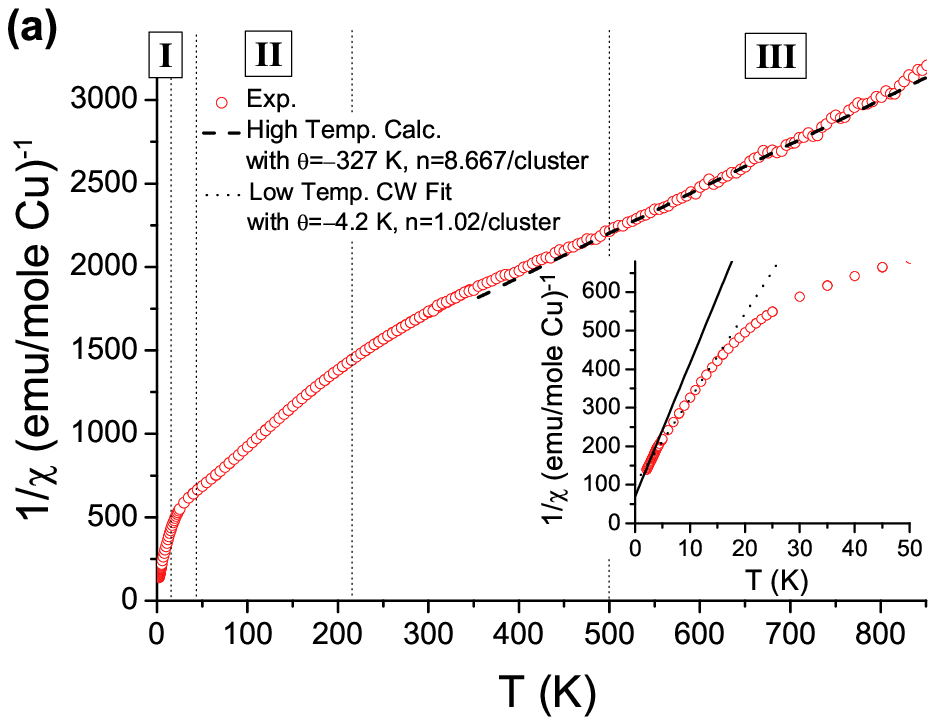}
\end{minipage}
\hspace{-0.2cm}
\begin{minipage}[b]{0.5\linewidth}
\centering
\includegraphics[scale=0.7]{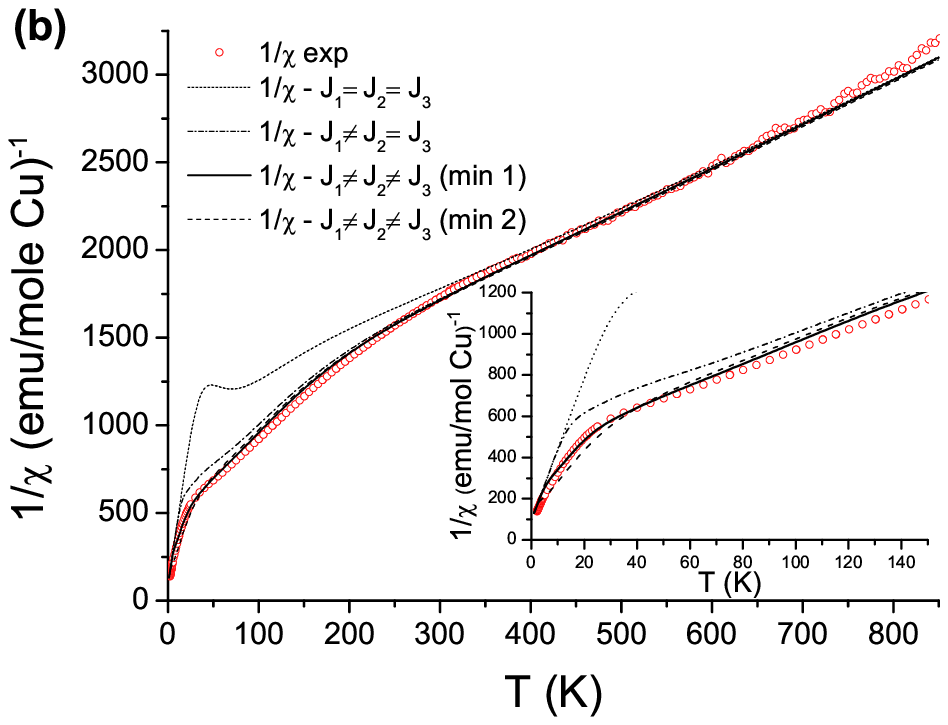}
\end{minipage}
\caption{(a) Inverse of the linear magnetic susceptibility of
La$_3$Cu$_2$VO$_9$ (hollow circles) measured in 0.1 T. The dash line
results from a Curie-Weiss fit at high temperature. The vertical
dotted lines materialize the 3 distinguished regions of quasi linear
magnetic behaviors. The inset shows details of the thermal variation
in region (I) with the forced Curie-Weiss fit (dotted line) and the
expected behavior for an ensemble of 2/3 9-Cu clusters and 1/3 8-Cu
clusters (continuous line) (b) Comparison of the inverse of the
measured susceptibility (circles) and of the calculated one (lines)
from different models of intra-cluster interactions taking into
account 8-Cu and 9-Cu clusters: $J_1=J_2=J_3$, $J_1 \neq J_2=J_3$,
$J_1 \neq J_2 \neq J_3$ (see fig. \ref{fig:substitutions} (b) and
sec. \ref{sec:model}).} \label{fig:invX}
\end{figure}

Magnetization measurements were performed at the Laboratoire Louis
N\'eel on a commercial Quantum Design MPMS SQUID magnetometer,
from 2~K to 350~K under a magnetic field up to 5~T, and by the
axial extraction method on purpose built magnetometers, which are
less sensitive but allow measurements in wider ranges of
temperature, up to 800~K, and of magnetic field, up to 10~T.

The isothermal magnetization curves, recorded at various
temperatures, do not show any obvious signature of a transition
towards a long range magnetic order, although there may be a slight
shape difference between the magnetic isotherms at 1.6~K and 3~K
\cite{robert}.
The magnetic linear susceptibility $\chi=M/H$ was recorded under a
0.1~T magnetic field where the isothermal magnetization is still
linear. Its inverse is shown as a function of the temperature in
Fig. \ref{fig:invX} (a). This last curve has a peculiar shape, made
of three quasi-linear regions at low (I), medium (II) and high (III)
temperatures with different slopes.
It was earlier suggested \cite{vander} that these would emerge
from the paramagnetic Curie-Weiss behavior $\chi=C/(T-\theta)$ of
distinct magnetic entities interacting with different interactions
(accounted for by the Curie-Weiss temperature $\theta$).
$C=ng^2\mu_BS(S+1)/3k_B$ ($g$=2) was evaluated per cluster with
$n$ the number of magnetic entities in the cluster and $S$ their
spin value.

\begin{table}
\begin{center}
\begin{tabular}{ccccc}
\hline Region & T range (K) & $\theta$ (K) & $\mu_{eff}$ $(\mu_B)$ &
$n$ \\ \hline
I & 2-11 & $-4.2 \pm 1.1$ & 1.732 & $1.02 \pm 0.19$ \\
III & 500-850 & $-327 \pm 73$ & 1.732 & 8.667
\\ \hline
\end{tabular}
\end{center}
\caption{Curie-Weiss parameters resulting from the fit
of the inverse susceptibility in the two quasi-linear regimes at low
and high temperature.} \label{tab:CW}
\end{table}

In a preliminary step we adopted the same analysis. We show in
figure \ref{fig:invX} (a) the results of our analysis. The
numerical values of $n$, $\theta$, the effective moment
$\mu_{eff}=g\sqrt{S(S+1)}$, and the temperature range of the fit
corresponding to the linear portion of regions (I) and (III) are
all reported in Table \ref{tab:CW}.
Assuming $S=1/2$, the slope of the high temperature region (III) is
well described above 500 K by the Curie-Weiss behavior with $n=
8.667$ ($=2/3 \times 9 + 1/3 \times 8$), which takes account of the
Cu/V substitution. The resulting large negative Curie-Weiss
temperature for this regime, $\theta=-327$ K, denotes a strong
intra-cluster antiferromagnetic coupling between these spins.

Concerning the intermediate temperature region (II), the
Curie-Weiss analysis leads to $n\simeq 4$ spins $1/2$ per cluster.
In section~\ref{sec:model}, our analysis through exact solution of
the cluster hamiltonian shows however that this does not have any
physical meaning since no collective magnetic entities are
actually formed \cite{robert}.

Unlike in ref. \cite{vander}, we found it more difficult to
isolate a linear regime in $1/\chi$, which rather presents a
continuous curvature (inset of Fig. \ref{fig:invX} (a)). A forced
Curie-Weiss fit in the reduced temperature range below 15 K yields
approximately 1 spin $S=1/2$ per cluster, which would suggest that
the magnetic entities in the low temperature range are collective
pseudo-spins, resulting from the entanglement of the wave
functions of the paramagnetic spins within each cluster at high
temperature. Since $\theta=-4.2$ K, these pseudo-spins would be
weakly antiferromagnetically coupled to each other. We notice that
this $\theta$ value is by two orders of magnitude smaller than
that in region (III), namely inter-cluster coupling is much weaker
than intra-cluster coupling. A main trouble with this low
temperature analysis is that $n=1.02$, which is inconsistent with
the existence of 8-Cu clusters for which a collective pseudo-spin
$S=0$ would be expected. We notice however that the slope of
$1/\chi$ increases, i.e $n$ decreases, when the temperature
decreases, which would suggest the existence of thermally
populated low energy magnetic excitations for 8-Cu clusters. This
requires a more stringent analysis of the magnetism of the
clusters through exact computation of model hamiltonians.
%

\section{\label{sec:model} Hamiltonian analysis}

Each plane of \lacuvo~is a triangular lattice of 9-spins clusters
with localized spins 1/2 on each Cu coupled through superexchange
via interstitial oxygens.
At first glance, the experimental study reveals that two sets of
antiferromagnetic exchange interactions are effective in this
system. The strongest one couples Cu$^{2+}$ ions within a cluster,
while the weakest one couples clusters together.
From the susceptibility measurements, the latter is two orders of
magnitude smaller.
As a consequence, there are two scales of frustration.
We shall now focus on the single cluster physics and leave for a
forthcoming publication \cite{robert} the inter cluster coupling
as the magnetic properties of \lacuvo\ should be dominated by that
of isolated clusters, at least for $T~\gtrsim$~2~K. However when
comparing the calculation to the experiment, the inter-cluster
coupling is taken into account within the molecular field
approximation.

In a single cluster approach, the calculation of the \lacuvo\
magnetic properties are obtained through exact diagonalisation of
the hamiltonian matrix of the modelized system. We shall always
consider a statistical ensemble of $2/3$ 9-Cu clusters and $1/3$
8-Cu clusters. At first, the clusters are all described by the
Heisenberg hamiltonian
\begin{eqnarray}
\label{eq:heis} \mathcal{H}=- J \sum_{ \langle ij \rangle}
\mathbf{S}_i \cdot \mathbf{S}_j
\end{eqnarray}
\noindent with the same antiferromagnetic exchange constant $J < 0$
limited to nearest neighbor spins $\mathbf{S}_i$ and $\mathbf{S}_j$
(Fig.~\ref{fig:substitutions}).
As shown in Fig.~\ref{fig:invX} (b), the different experimental
regimes of the inverse magnetic susceptibility, fingerprint of the
collective pseudo-spin formation, are not correctly reproduced in
the calculation, pointing out the limitation of this simple cluster
model. On scaling the calculated $\chi=2/3 \chi_9 + 1/3 \chi_8$,
resulting from the weighted average of the 8-spins, $\chi_8$, and
9-spins, $\chi_9$, susceptibilities, to the high temperature regime
(III), we get $J \simeq -385$ K as a crude estimation of the
exchange interaction.

As to improve the model, the distortion of the 8-Cu clusters and the
subsequent change in the exchange interactions accompanying the Cu/V
substitution, evidenced in the DFT calculations, was taken into
account by considering three different exchange parameters as
depicted in figure \ref{fig:substitutions} (b).
The susceptibility has been calculated, by varying systematically
the three exchange interaction constants and comparing results to
the experimental inverse susceptibility, imposing first $J_2=J_3$,
then leaving these to vary freely with respect to each other. When
$J_2=J_3$, the best agreement is obtained for
$\{J_1,J_2,J_3\}=\{1.06,0.27,0.27\} J$ with $J=-385$ K, (see Fig.
\ref{fig:invX} (b)), while for $J_2\neq J_3$, two minima in the
goodness of fits were found in the $\{J_1,J_2,J_3\}$ parameter
space: $\{J_1,J_2,J_3\}=\{1.06,0.42,0.05\} J$ (minimum 1) and
$\{J_1,J_2,J_3\}=\{1.06,0.05,0.2\} J$ (minimum 2) with $J=-385$ K,
(see Fig. \ref{fig:invX} (b)).
The model with $J_2\neq J_3$ is an improvement over that with $J_2
= J_3$. It leads to a better agreement with the low temperature
experimental behavior (region (I)), in particular by providing a
curved thermal variation of the inverse of the magnetic
susceptibility. On examining the structure of the hamiltonian
spectrum, this is explained through as due to the existence of
magnetic excitations lying close to the collective $S=0$ state on
the 8-Cu clusters.

Actually, there is no reason to assume that the exchange in 9-Cu
clusters are identical. Nevertheless, correct adequation to
experiment is obtained within the $\{J_1,J_2,J_3\}$ model that
takes into account only the effect of Cu/V substitutions. This
suggests that the real exchange interactions couplings in the 9-Cu
clusters should not differ that much from the uniform one. A
detailed examination of its energy spectrum confirms that a
collective pseudo-spin $S=1/2$ does stabilize in the 9-Cu clusters
at temperatures below 15 K. On computing the quantum statistical
average of the square modulus of the collective spin, we also
confirm that in region (II), a Curie-Weiss analysis has no meaning
\cite{robert}.

\section{Conclusion}

The \lacuvo\ oxide compound is constituted of 8-spins and 9-spins
clusters, laid out on 4 vertex-sharing triangles, basing block of
the kagom\'e lattice. From comparison of magnetization
measurements and exact diagonalisation of the Hamiltonian of a
spins cluster models, the low temperature stabilization on each
9-spins cluster of a collective pseudo-spin $S$=1/2 resulting from
the entanglement of the spin wave functions of the 9 paramagnetic
spins 1/2 at high temperature, is evidenced.
This opens the possibility to study the gradual coupling of these
pseudo spins 1/2 at low temperature from the experimental as well
as from the theoretical point of view and provides a unique
experimental realization of dimensional crossover tuned by temperature.


\end{document}